\documentclass[12pt]{article}

\usepackage{graphicx,a4,epsfig,rotate,amsmath}    
\usepackage{latexsym,color,citesort,amsfonts}
\newcommand{\eq}{\begin{eqnarray}}
\newcommand{\en}{\end{eqnarray}}

\newcommand{\ed}{\end{document}}
\newcommand{\bc}{\begin{center}}
\newcommand{\ec}{\end{center}}

\begin{document}

\thispagestyle{empty}

\begin{flushright}
{\tiny{FZJ-IKP-TH-2009-34}}
{\tiny{HISKP-TH-09/35}}
\end{flushright}

\vspace{1.2cm}
\bc{\Large{\bf Resonances in an external field:\\[2mm] the 1+1 dimensional case }}\footnote{
Work supported in part by DFG (SFB/TR 16,
``Subnuclear Structure of Matter'') and  by the Helmholtz Association
through funds provided to the virtual institute ``Spin and strong
QCD'' (VH-VI-231). We also acknowledge the support of the European
Community-Research Infrastructure Integrating Activity ``Study of
Strongly Interacting Matter'' (acronym HadronPhysics2, Grant
Agreement n. 227431) under the Seventh Framework Programme of EU.
A.R. acknowledges financial support 
of the Georgia National Science Foundation (Grant \#GNSF/ST08/4-401).}

\vspace{0.5cm}
11 January 2010

\vspace{0.5cm}

D.~Hoja$^a$,
U.-G.~Mei{\ss}ner$^{a,b}$ and  
A.~Rusetsky$^a$

\vspace{2em}

\begin{tabular}{c}
$^a\,$Universit\"at Bonn, 
Helmholtz--Institut f\"ur Strahlen-- und Kernphysik (Th)\\ and
Bethe Center for Theoretical Physics,
D-53115 Bonn, Germany\\[2mm]
$^b\,$Forschungszentrum J\"ulich, Institut f\"ur Kernphysik 
(IKP-3),\\ J\"ulich Center for Hadron Physics\\
and Institute for Advanced Simulation (IAS-4), D-52425 J\"ulich, Germany\\
\end{tabular} 

\ec

\vspace{1cm}

{\abstract{
\noindent
Using non-relativistic effective field theory in 1+1 dimensions, we
generalize L\"u\-scher's approach for  resonances in the presence
of an external field. 
This generalized approach provides a framework to
study the infinite-volume limit of the form factor
of a resonance determined in lattice simulations.  
}}

\vskip1cm

{\footnotesize{\begin{tabular}{lll}
{\bf{Pacs:}}$\!\!\!\!$&\quad&  11.15.Ha, 12.38.Gc, 11.10.St\\
{\bf{Keywords:}}$\!\!\!\!$&\quad& Lattice field theory, resonances, finite volume,
L\"uscher formula, magnetic moment

\end{tabular}}
}
\clearpage

\section{Introduction}

Recently, the external field method has been widely used 
to calculate  magnetic moments and polarizabilities of both  stable
and excited hadrons
in lattice QCD~\cite{Bernard:1982yu,Lee:2005ds,Lee:2008qf,Aubin:2008qp,Christensen:2004ca,Lee:2005dq,Detmold:2009dx,Tiburzi}. In an alternative approach,
the three-point function, evaluated in lattice QCD, is
extrapolated in the photon momentum squared variable $q^2$
to the value $q^2=0$ (see, e.g.,~\cite{Alexandrou:2007we,Alexandrou:2008bn}).
Both methods are justified in case of stable particles. In case of
excited hadron states, however,  conceptual problems arise. As it
is well known,  resonances are not described by a single level in the 
spectrum of the Hamiltonian. Rather, they characterize the whole spectrum and
are usually extracted by applying L\"uscher's formula 
to the lattice data at different lattice volumes~\cite{Luescher}.
In order to proceed with the determination of, e.g. the magnetic moment
of a resonance, the generalization of L\"uscher's approach in the presence
of external fields is needed. For example, it is not clear, how the real
and discrete 
energy eigenvalues of a system, placed in an external field, are related 
to the resonance form factor, which is a complex quantity in general.
To the best of our knowledge, such a generalization
is not available at present and we will fill this gap in the
following.

In order to extract the 
form factors of resonances on the lattice, one has to
thoroughly study the volume dependence of these quantities. As it is well 
known,  non-relativistic effective field theories (NR EFTs) provide the most
convenient framework to systematically address this problem. In the past
such kind of theories have been used to obtain a simple derivation of L\"uscher's 
formula~\cite{Beane}, including the cases of particles with spin~\cite{Lage}
and multi-channel scattering~\cite{Lage:2009zv}. Moreover, EFT techniques have been
applied to the calculation of the volume dependence of the meson and
nucleon form factors~\cite{Hu:2007eb,Tiburzi:2007ep,Tiburzi:2008pa}. 
In this paper, using an  effective field theory in 1+1 dimensions, we discuss
the generalization of L\"uscher's approach in the presence of an external 
field. The outline of the procedure is as follows. We first assume that, 
in the absence of the external field, the 
standard effective-range expansion converges in a part of the complex momentum
plane including the (complex) resonance pole denoted by $p_*$. 
This fact can be used to perform
the analytic continuation of L\"uscher's formula and to find the position
of the pole. The key observation is that the limit $p\to p_*$ in L\"uscher's
formula corresponds to the infinite-volume limit.
In turn, this leads to the conclusion that the shift of the
pole position in the external field, evaluated at lowest order in this field,
is proportional to the resonance form factor evaluated in the infinite volume
and on the mass shell $p\to p_*$.

The situation in 3+1 dimensions is different from 1+1 dimensions in one 
important aspect. Namely, 
whereas in 1+1 dimensions the limit $p\to p_*$ {\em always} implies
the infinite-volume limit, we expect that 
there are in addition what we term as 
{\em finite fixed points} in 3+1 dimensions. Owing to this fact, 
together with some additional technical complications which arise in the
3+1-dimensional case, we find it convenient to separate these two cases.
In the present paper, we limit ourselves to a toy model of a resonance
in an external field in 1+1 dimensions. The 3+1-dimensional case 
will be addressed in future publications. 

The outline of the paper is as follows. In section~\ref{sec:without}
we construct the non-relativistic effective theory, introducing the resonance
field as an independent degree of freedom. Fixing
 the pole position in the two-point function of the resonance field in the
complex plane is discussed. Turning on the external field and the resulting
shift of the pole position is considered in section~\ref{sec:with}.
The relation of the mass-shell limit $p\to p_*$ and the infinite-volume 
limit
is discussed in section~\ref{sec:pstar}. In section~\ref{sec:threepoint}
we address the evaluation of the three-point function related to the form
factor under consideration. Section~\ref{sec:concl} contains our conclusions.
Finally, the calculation of the loop function in the infinite and in a 
finite volume is relegated to the appendix~\ref{app:Z00}.

\section{Non-relativistic field theory in the absence of an external field}
\label{sec:without}

Below, we shall use the ``relativized'' version of NR EFT, which has
been first introduced in Ref.~\cite{CGKR}. The theory in a finite volume was
first considered in Refs.~\cite{Lage,Lage:2009zv}. For all details, we refer the
interested reader to these articles. Here, we 
consider two massive scalar non-relativistic fields $\Psi$ and $\Phi$ in
 1+1 dimensions. The non-relativistic Lagrangian that describes these 
particles is given by
\eq\label{eq:L}
 {\cal L}&=&\Psi^\dagger \left(i\partial_t-w_\Psi(\nabla^2)\right) 2w_\Psi(\nabla^2)\Psi 
+\Phi^\dagger \left(i\partial_t-w_\Phi(\nabla^2)\right) 2w_\Phi(\nabla^2)\Phi
\nonumber\\[2mm]
&+&C_0\Psi^\dagger\Phi^\dagger\Phi\Psi
+C_1\biggl\{(\Psi^\dagger\stackrel{\leftrightarrow}{\nabla^2}\Phi^\dagger)\,
\Phi\Psi+\text{h.c.}\biggr\}+\cdots\, ,
\en
where $w_i(\nabla^2)\doteq (m_i^2-\nabla^2)^{1/2}$ for  $i=\Psi,\Phi$, and
$(u\!\stackrel{\leftrightarrow}{\nabla^2}\!v)\doteq u\nabla^2v+v\nabla^2u$. 
The low-energy constants $C_0,C_1,\ldots$ are related to the effective-range
expansion of the $\Psi\Phi$ elastic scattering amplitude and the ellipses 
stand for the terms with higher derivatives. At tree level,
this scattering amplitude in the center-of-mass frame is given by the expression
\eq\label{eq:tree}
H(p,q)=C_0-2C_1(p^2+q^2)+\cdots \, ,
\en
where $p,q$ denote the relative momenta of the $\Psi\Phi$-pair in the 
final and initial state, respectively.

In the non-relativistic field theory, 
the full scattering amplitude $T(p,q)$ is obtained by summing up the tree-level
result given in Eq.~(\ref{eq:tree}) to all orders. 
The result for the on-shell amplitude
in the CM frame is given by 
\eq\label{eq:Kmatr}
 T(p,p)=\frac{H(p,p)}{1-J(p)H(p,p)}\, ,
\en
where $J(p)$ denotes the $\Psi\Phi$-loop integral, evaluated by expanding 
the integrand in powers of momenta and using dimensional regularization 
(see Ref.~\cite{CGKR} and Appendix~\ref{app:Z00} for more details), in the center-of-mass frame we obtain 
\eq\label{eq:Jp}
 J(p)&=&\int\frac{d^Dk}{(2\pi)^Di}\frac{1}{2w_\Psi(k)2w_\Phi(k)}
\frac{1}{\left(w_\Psi(k)-P^0+k^0\right)\left(w_\Phi(k)-k^0\right)}\nonumber\\[2mm]
&=&
\frac{i}{4P^0p}+O(D-2)\, ,
\en
where
$P^0=w_\Psi(p)+w_\Phi(p)$ is the center-of-mass energy of the 
$\Psi\Phi$ system, and
 $p$ stands for the relative momentum of the $\Psi\Phi$-pair.

The unitarity relation is
\eq\label{eq:unit}
\mbox{Im}\,T(p,p)=\frac{1}{4P^0p}\,|T(p,p)|^2\, ,\quad\quad
T(p,p)=\frac{4P^0p}{\cot\,\delta(p)-i}\, ,
\en
where $\delta(p)$ is the scattering phase. Comparing Eqs.~(\ref{eq:Kmatr})
and (\ref{eq:unit}), we obtain
\eq\label{eq:effrange}
 p\cot\delta(p)=\frac{4P^0p^2}{H(p,p)}=A_1p^2+A_2p^4+\cdots\, ,
\en
where the effective range expansion parameters $A_1,A_2\ldots$ can be 
expressed in terms of the effective couplings $C_0,C_1,\ldots$, 
see Eq.~(\ref{eq:tree}).
Eq.~(\ref{eq:effrange})
 implies that the quantity  $p\cot\delta(p)$ is an analytic
function of the variable $p^2$ in the region of small $p^2$, where the 
effective range expansion converges.
If there is a low-lying resonance in the $\Psi\Phi$ elastic scattering,
the amplitude will have a pole on the second Riemann sheet. 
The pole position $p_*$ 
is determined by the equation $p_*\cot\delta(p_*)=-ip_*$, where
the analytic continuation to complex values of $p^2$
is performed by using the effective-range expansion displayed
in Eq.~(\ref{eq:effrange})~\footnote{Note the sign on 
the r.h.s. of this equation, which corresponds to the
choice of the second sheet.}.

Suppose there is a low-lying resonance in  $\Psi\Phi$ elastic scattering.
In this case, it is convenient to use the equivalent formulation of the non-relativistic
effective field theory, introducing an auxiliary  ``resonance field'' $\chi$.
The Lagrangian of this theory is given by
\eq\label{eq:Lchi}
{\cal L}_\chi&=&\Psi^\dagger \left(i\partial_t-w_\Psi(\nabla^2)\right) 
2w_\Psi(\nabla^2)\Psi
+\Phi^\dagger \left(i\partial_t-w_\Phi(\nabla^2)\right) 2w_\Phi(\nabla^2)\Phi
\nonumber\\[2mm]
&+&\chi^\dagger \left(i\partial_t-w_\chi(\nabla^2)\right) 
2w_\chi(\nabla^2)\chi
+f_0\left(\Psi^\dagger\Phi^\dagger\chi+\text{h.c.}\right)
\nonumber\\[2mm]
&+&f_1\left((\Psi^\dagger\stackrel{\leftrightarrow}{\nabla^2}\Phi^\dagger)\chi
+\text{h.c.} \right)\cdots \, ,
\en
where $f_0,f_1,\ldots$ are the low-energy couplings and $m_\chi$ in
$w_\chi(\nabla^2)=(m_\chi^2-\nabla^2)^{1/2}$ is a {\em real} parameter.
Note that we opted for elimination the 4-point vertices 
with $\Psi$ and $\Phi$ fields in the
Lagrangian. Since the ultraviolet divergences do not emerge in the dimensional
regularization, such 4-particle vertices are not generated by the loops either.

\begin{figure}[t]
\begin{center}
\includegraphics[width=14.cm]{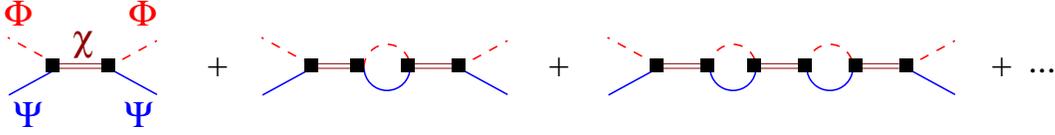}
\end{center}
\caption{The $\Psi\Phi$ elastic scattering amplitude in the theory described
by the Lagrangian ${\cal L}_\chi$ given in Eq.~(\ref{eq:Lchi}). Solid, dashed
and double lines denote the $\Psi,\Phi$ and $\chi$ fields, respectively. The 
shaded squares denote the full three-particle vertex $\gamma(p)$.}
\label{fig:bubblesum}
\end{figure}

The $\Psi\Phi$ elastic scattering amplitude in this framework is generated by
summing up the bubble diagrams shown in Fig.~\ref{fig:bubblesum}
\eq\label{eq:fullTchi}
T(p,p)=\gamma^2(p)S_\chi(p)\, ,\quad\quad
S_\chi(p)=\frac{1}{2m_\chi(m_\chi-P^0)-\Sigma_0(P^0)}\, ,
\en
where $S_\chi$ stands for the propagator of the $\chi$ particle,
$\Sigma_0(P^0)=\gamma^2(p)J(p)$ is the self-energy of the field $\chi$,  
and 
\eq
\gamma(p)=f_0-2f_1p^2+\ldots
\en
denotes the full $\Psi\Phi\chi$ 
vertex function.

Comparing Eqs.~(\ref{eq:Kmatr}) and (\ref{eq:fullTchi}) gives the 
matching between the two equivalent formulations
\eq\label{eq:matching}
H^{-1}(p,p)=\frac{2m_\chi(m_\chi-P^0)}{\gamma^2(p)}\, .
\en
Further, the relation between the couplings $C_i$ and $f_i$ can be 
obtained by expanding
both sides of the above equation in powers of $p^2$. In case when the 
low-lying resonance is present, the couplings $f_i$ are of natural size, 
whereas the couplings $C_i$ contain powers of the small scale
$\Delta_0=m_\chi-m_\Psi-m_\Phi$ in the denominator and may 
become unnaturally large. The position of the resonance pole is given by
\eq\label{eq:poleposition}
\frac{8m_\chi P^0_*p_*^2(m_\chi-P^0_*)}{\gamma^2(p_*)}=-ip_* \, ,
\en
where $P^0_*=w_\Psi(p_*)+w_\Phi(p_*)$.

In order to obtain
the energy spectrum of the system described by the Lagrangian~(\ref{eq:Lchi}),
placed in a box of size $L$, we consider the finite-volume propagator
of the particle $\chi$
\eq\label{eq:SL}
S^L_\chi(p)=\frac{1}{2m_\chi(m_\chi-P^0)-\gamma^2(p)J^L(p)}\, ,
\en
where
\eq\label{eq:JL}
&&\hspace*{-.63cm} J^L(p)=\frac{1}{L}\,\sum_k
\int\frac{dk^0}{2\pi i}\frac{1}{2w_\Psi(k)2w_\Phi(k)}
\frac{1}{\left(w_\Psi(k)-P^0+k^0\right)\left(w_\Phi(k)-k^0\right)}\, ,
\nonumber\\
&&
\en
and $k$ denotes the discrete lattice momenta 
$k=2\pi n/L\, ,\,\, n\in \mathbb{Z}$.
 
It can be shown (see Appendix~\ref{app:Z00}) that, up to the terms that
vanish exponentially with $L$, the function $J^L(p)$ is given by
\eq\label{eq:Z00}
J^L(p)=-\frac{1}{4P^0p}\,\cot\,\frac{pL}{2}\, .
\en
Using Eqs.~(\ref{eq:effrange}), (\ref{eq:matching}) and (\ref{eq:Z00}),
it can be shown that the denominator in Eq.~(\ref{eq:SL}) vanishes for
the values of $p$ which obey the relation $\cot\,\delta(p)=
-\cot\,(\frac{1}{2}\,pL)$. 
This leads to the  L\"uscher equation~\cite{Luescher} in 1+1 dimensions
(we take the positive root $p>0$)
\eq
2\delta(p)=-pL+2\pi m\, ,\quad\quad m\in \mathbb{N}\, .
\en
Let us restrict ourselves to a fixed energy level ($m$ fixed)
and measure $p_m=p_m(L)$ for different values of $L$ on the lattice.
Substituting this into L\"uscher's equation, we get 
\eq\label{eq:Lmp}
p\,\cot\,\delta_m(p)=
-p\,\cot\,(\frac{1}{2}\,pL_m(p))\, ,
\en
where the function $L=L_m(p)$ is the 
inverse of $p=p_m(L)$ and the index $m$ attached to the scattering phase
indicates that it has been determined from the fixed level $m$. 

The left-hand side of Eq.~(\ref{eq:Lmp}) is a function of the variable
$p$ only. Fitting this function, extracted from the lattice measurement,
to the effective range expansion in Eq.~(\ref{eq:effrange}), one may determine
the {\em real} coefficients $A_1^m,A_2^m,\ldots$ (the index $m$ again indicates
the fixed energy level). If the volume $L$ is large enough (as required 
in the derivation of L\"uscher's equation), these coefficients do  indeed not
depend on the index $m$, up to  exponentially suppressed terms.

Finally, assuming that the effective range expansion is convergent in the range
of $p^2$ we are working, one may solve the equation $A_1p_*^2+A_2p_*^4+\ldots=
-ip_*$ to obtain the complex pole position on the second Riemann sheet.
The pole position remains stable up to exponentially
suppressed terms (for this reason,
we suppress the index $m$ in the quantities $A_i$ here).

Suppose now that we consider the limit $p\to p_*$ in Eq.~(\ref{eq:Lmp}), which
we rewrite in the following form (the index $m$ is suppressed)
\eq\label{eq:topole}
\cot\,\delta(p)=\frac{A_1p^2+A_2p^4\cdots}{p}=-\cot\,\pi q\, ,
\quad\quad
q=\frac{pL(p)}{2\pi}~,
\en
where the {\em known} coefficients $A_1,A_2,\ldots$ are assumed to be extracted from lattice
measurements at {\em real} values of $p$.
At the pole position $p\to p_*$, the left-hand side of Eq.~(\ref{eq:topole})
tends to $-i$. Then, as it can be easily seen,
$\text{Im}\,q\to -\infty$. The real part of the variable $q$ stays finite
in this limit and depends on the path along which the pole  $p_*$ is approached.
The real and imaginary parts of the ``volume'' $L(p)$,
defined by Eq.~(\ref{eq:topole}), also diverge in this limit.

The above observation plays the central role in study of the properties of a
resonance in the external field. Loosely spoken, it states that the 
mass-shell limit for a resonance implies the infinite-volume limit. For
this reason, e.g. the shift of the resonance pole position in an external 
field is determined by the vertex function at zero momentum transfer, 
calculated in the infinite volume.

\section{Turning on the external field}
\label{sec:with}

In order to study the behavior of the system placed in an external field,
one has to equip the Lagrangian with that field. 
Instead of introducing an electromagnetic field in two dimensions, 
we consider a toy model with a 
constant scalar external field $v$. To ease the notation,
we in addition assume that the field $v$ does not couple to $\Phi$.

The modified Lagrangian takes the form
\eq
{\cal L}^v_\chi={\cal L}_\chi+v\biggl(\Psi^\dagger\Psi+\lambda_1\chi^\dagger\chi
+\lambda_2(\Psi^\dagger\Phi^\dagger\chi+\text{h.c.})\biggr)\, ,
\en
where the couplings $\lambda_i$ are dimensionless,
It should be pointed out that, since the UV divergences are absent 
in dimensional regularization, the terms with derivatives acting on the 
fields, are not generated by loop corrections. Hence, one may omit these 
altogether without loss of generality.

\begin{figure}[t]
\begin{center}
\includegraphics[width=12.cm]{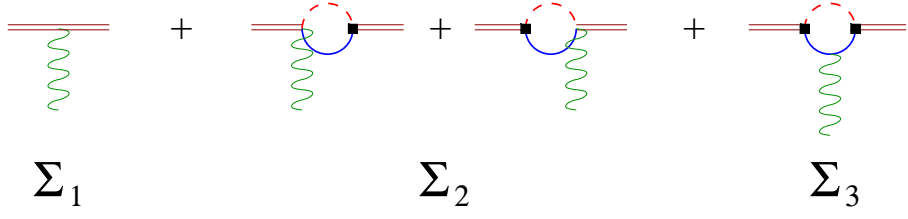}
\end{center}
\caption{Diagrams contributing to the self-energy of the $\chi$ field at order 
$v$. Solid, dashed and double lines denote $\Psi,\Phi$ and $\chi$, 
respectively,
and the wiggly line stands for the external field. The shaded squares
denote the full three-particle vertex $\gamma(p)$.}
\label{fig:vertex}
\end{figure}

Let us first restrict ourselves to the infinite volume.
In the presence of the external field, the pole position in the two-particle function
of the field $\Psi$ shifts from $P^0=(m_\Psi^2+p^2)^{1/2}$
to $P^0=(m_\Psi^2-v+p^2)^{1/2}+O(v^2)=w_\Psi(p)-v/(2w_\Psi(p))+O(v^2)$, whereas the $\Phi$-pole stays put. 
Further, the shift of the pole position in the 
elastic $\Psi\Phi$ scattering amplitude to first order in the external 
field $v$ is equal to the shift in the two-point function
of the auxiliary field $\chi$ and can be determined by solving the following 
equation:
\eq
(S_\chi^v(p))^{-1}=2m_\chi(m_\chi-P^0)-\Sigma_0-v(\Sigma_1+\Sigma_2+\Sigma_3)+O(v^2)=0\, ,
\en
where $\Sigma_{1,2,3}$ denote the different contributions to the self-energy of 
the $\chi$-particle at order $v$, which are shown in Fig.~\ref{fig:vertex}.

The (complex) shift of the pole position to this order is given by
\eq\label{eq:poleshift}
\delta P_*^0=-\frac{v(\Sigma_1(P_*^0)+\Sigma_2(P_*^0)+\Sigma_3(P_*^0))}
{2m_\chi+\Sigma_0'(P_*^0)}+O(v^2)\, ,
\en
where $\Sigma_0'(P^0)\doteq\frac{d}{dP^0}\,\Sigma_0(P^0)$. 
Further, using Eq.~(\ref{eq:fullTchi}), for the wave function renormalization
factor of the $\chi$ particle in the absence of the external field
one obtains $Z_\chi^{-1}=2m_\chi+\Sigma_0'(P_*^0)$. Consequently, the resonance
pole shift (\ref{eq:poleshift}) in the infinite volume is given by the
on-mass-shell vertex function of $\chi$, evaluated at zero momentum transfer
\eq
\delta P_*^0=-v\Gamma(0,0)\, ,\quad\quad
\Gamma(0,0) = Z_\chi\, \left(\Sigma_1(P_*^0)+\Sigma_2(P_*^0)+\Sigma_3(P_*^0)\right)\, .
\en
The quantity $\Gamma(0,0)$ is the analog of the magnetic moment of a resonance
in our toy model\footnote{For the discussion of the magnetic moment of
unstable particles, see, e.g.,~\cite{Gegelia:2009py} and references therein.}.

Next, we consider the same system in a finite volume. The energy spectrum
at $O(v)$ is obtained by solving the equation
\eq\label{eq:spectrum}
2m_\chi(m_\chi-P^0)-\Sigma_0^L-v(\Sigma_1^L+\Sigma_2^L+\Sigma_3^L)=0\, ,
\en
where $\Sigma_{0,1,2,3}^L$ are defined by the same expressions as in
the infinite volume, where the integration over the loop momentum $k$ is
replaced by a discrete sum, see, e.g., Eq.~(\ref{eq:JL}).

Suppose that $P^0_m,~m\in\mathbb{N}$, is the spectrum at $v=0$.
Turning on the external field leads to a shift $P^0_m\to P^0_m+\delta P^0_m$,
where $\delta P^0_m=O(v)$. From Eq.~(\ref{eq:spectrum}) one obtains a 
relation, which is very similar to Eq.~(\ref{eq:poleshift}) but contains only
{\em real} variables
\eq\label{eq:poleshiftR}
\delta P_m^0=-\frac{v(\Sigma_1^L(P_m^0)+\Sigma_2^L(P_m^0)+\Sigma_3^L(P_m^0))}
{2m_\chi+(\Sigma_0^L(P_m^0))'}+O(v^2)\, ,
\en
where
\eq
\Sigma_0^L(P^0)&=&\gamma^2(p)J^L(p)\, ,
\quad
\Sigma_1^L(P^0)=\lambda_1\, ,
\quad
\Sigma_2^L(P^0)=2\lambda_2\gamma(p)J^L(p)\, ,
\nonumber\\[2mm]
\Sigma_3^L(P^0)&=&\frac{\gamma^2(p)}{2w_\Psi(p)}\,
\frac{d}{dP^0}\,J^L(p)\biggr|_{L=\text{const}}\, ,
\en
with $J^L(p)$ given by Eq.~(\ref{eq:Z00}).

One may easily perform the limit $L\to \infty$ in Eq.~(\ref{eq:poleshiftR}),
using $\gamma^2(p_m)J^L(p_m)=2m_\chi(m_\chi-P^0_m)$ 
(where $p_m$ is defined through $P_m^0=w_\Psi(p_m)+w_\Phi(p_m)$)
and
the fact that, {\em for a fixed index $m$}, $P_m^0\to m_\Psi+m_\Phi$ and
$p_m\to 0$, as $L\to \infty$. The result is given by
\eq\label{eq:impulse}
\lim_{L\to\infty}\delta P_m^0=-\frac{v}{2w_\Psi(p_m)}+O(v^2)\, .
\en
The physical interpretation of the above result is very transparent.
Indeed, Eq.~(\ref{eq:impulse}) exactly reproduces the result obtained in the
impulse approximation. This means that a level with any fixed index $m$ disintegrates
in the limit $L\to\infty$, and the magnetic moment of a system is given by a
sum of the magnetic moments of the constituents $\Psi$ and $\Phi$. In this
connection,  note that only $\Sigma_3$ contributes 
to Eq.~(\ref{eq:impulse}) in the limit $L\to\infty$
(e.g., the r.h.s. of this equation does not depend on the couplings 
$\lambda_{1,2}$).
This result, however, {\em clearly does not coincide} with
the known answer for the magnetic moment in the infinite-volume limit,
given by Eq.~(\ref{eq:poleshift}).

On the other hand, let us try extract the pole position on the second Riemann
sheet, following exactly the same path as in case with no external field.
Namely, measuring the volume-dependent energy spectrum on the lattice 
$p=p_m^v(L)$ and inverting this relation $L=L_m^v(p)$, 
one may parameterize the quantity
\eq\label{eq:effrange-v}
p\,\cot\biggl(\frac{pL_m^v(p)}{2}\biggr)=
\frac{4P^0p^2}{\gamma^2(p)}\,\biggl(-2m_\chi(m_\chi-P^0)
+v(\Sigma_1^L+\Sigma_2^L+\Sigma_3^L)\biggr)
\en
as a certain known function of a real variable $p$ (see next section for the details). 
As the next step, one uses this parameterization to continue the expression
(\ref{eq:effrange-v}) into the complex plane and finds the shift of
the pole position. The result is given by
\eq\label{eq:poleshift-right}
\delta P_*^0=-\frac{v(\Sigma_1^L(P_*^0)+\Sigma_2^L(P_*^0)+\Sigma_3^L(P_*^0))}
{2m_\chi+\Sigma_0'(P_*^0)}+O(v^2)\, .
\en
Note that the self-energy operator calculated in the infinite volume, 
appears in the denominator.
Comparing Eqs.~(\ref{eq:poleshift}) and (\ref{eq:poleshift-right}), we see
that, in order to prove the equivalence of these two expressions at 
large volumes, one has
to investigate -- in a finite volume --
the analytic continuation of the diagrams shown
in Fig.~\ref{fig:vertex} into the complex plane.

\section{Analytic continuation}
\label{sec:pstar}

In this section we shall study the analytic continuation of the quantity,
which for real values of $p$ is given by Eq.~(\ref{eq:effrange-v}).
The self-energy in the presence of the external field consists of three
contributions $\Sigma_{1,2,3}^L$. The analytic continuation of the
first contribution is trivial because it is a constant 
$\Sigma_1^L(P^0)=\lambda_1$. 

The second contribution can be rewritten as
\eq\label{eq:Sigma2}
\Sigma_2^L(P^0)=2\lambda_2\gamma(p)J^L(p)
=\frac{4\lambda_2 m_\chi(m_\chi-P^0)}
{\gamma(p)}\, .
\en
It is immediately seen that on the trajectory $L=L_m(p)$ the quantity 
$\Sigma_2(P^0)$ does not depend on the index $m$ (up to exponentially 
suppressed terms) and is a 
low-energy polynomial in the variable $p^2$: 
$\Sigma_2^L(P^0)=B_0+B_1p^2+\ldots$. The analytic continuation into the complex
plane is achieved by substituting $p^2\to p_*^2$.
Using Eq.~(\ref{eq:poleposition}), it can be easily shown that
on the second Riemann sheet
\eq
\lim_{p\to p_*}\Sigma_2^L(P^0)=\frac{-i\lambda_2\gamma(p_*)}{2P^0_*p_*}
=2\lambda_2\gamma(p_*)J(p_*)=\Sigma_2(P_*^0)\, ,
\en
i.e., the mass-shell limit $p\to p_*$ implies the infinite-volume limit
$\Sigma_2^L(P^0)\to \Sigma_2(P_*^0)$.

The situation is more complicated for the last contribution 
\eq\label{eq:Sigma3}
\Sigma_3^L(P^0)=\frac{\gamma^2(p)w_\Phi(p)}{8p^3(P^0)^2}\,\biggl(
\biggl(1+\frac{p^2}{w_\psi(p)w_\phi(p)}\biggr)\cot\,\pi q_m
+\pi q_m\,(1+\cot^2\pi q_m)\biggr)\, ,
\en
where $q_m=pL_m(p)/(2\pi)$.
The quantity $\cot\,\pi q_m=-p(A_1+A_2p^2+\ldots)$ is proportional to a 
low-energy polynomial and can be straightforwardly continued in the complex
plane. It does not depend on the index $m$. The difficulty comes from the last
term which contains the function 
\eq\label{eq:Lm}
q_m(p)=-\frac{1}{\pi}(\delta(p)-\pi m)
=m-\frac{1}{2}+\frac{1}{\pi}\,
\text{arccot}(\cot\,\delta(p))\, .
\en 
As we see, this expression depends on the index $m$.

The inclusion of $\Sigma_3^L(P^0)$ modifies the 
effective-range expansion in a given constant external field. 
Namely the expansion of the cotangent of the phase shift
now starts at $p^{-2}$, depends on the level index $m$ and contains both odd and even powers in $p$.
Note, however, that the dangerous terms are multiplied by the factor 
$(1+\cot^2\pi q_m)$ which vanishes at the pole after analytic 
continuation. Consequently, the dependence on the index $m$ in the expression 
of the pole shift indeed disappears. Note that, in the analytic continuation
only the last term in Eq.~(\ref{eq:Lm}) is potentially problematic, 
because the arctangent
diverges at the pole. However, the factor $(1+\cot^2\pi q_m)$ 
plays the decisive role. It can be checked that
the function $f(z)=(1+z^2)\,\text{arccot}\,z$, albeit non-analytic at $z=\pm i$,
expanded in a Taylor series of $z$, converges to 0 at $z\to\pm i$
 (here, $z$ denotes $\cot\,\delta(p)$).

Finally, the procedure of the analytic continuation in the presence of the
external field can be formulated as follows. One measures the spectrum
$p=p_m^v(L)$ in the resonance region where $\cot\,\delta(p)$ is small
and inverts this relation $L=L_m^v(p)$. The function 
$\cot(\frac{1}{2}\,pL_m^v(p))$ in this region is 
fitted by the series
\eq 
\cot\biggl(\frac{1}{2}\,pL_m^v(p)\biggr)=\frac{D_{-2}}{p^2}+\frac{D_{-1}}{p}+D_{0}
+D_1p+D_2p^2+\cdots\, ,
\en
where the coefficients with even powers of $p$ may depend on the level index
$m$ (we recall that the series contains only odd powers of $p$ in the 
absence of the external field, see Eq.~(\ref{eq:topole})). 
Finally, after determining the coefficients $D_i$ from the fit,
we continue the result analytically by substituting $p\to p_*$ in this series\footnote{Moreover, since the dependence on the index $m$ in Eq.~(\ref{eq:Lm})
is linear, one may in principle
eliminate the whole contribution of the last term of Eq.~(\ref{eq:Sigma3}),
 e.g., by performing the measurements for different
energy levels.}.

\section{Matrix elements at zero momentum transfer}
\label{sec:threepoint}

An alternative technique for
evaluating the ``magnetic moment'' in our toy model is to
calculate the following Euclidean three-point function on the lattice
\eq\label{eq:G}
G(t',t)=\langle 0| \Delta(t') J(0) \Delta^\dagger(t)|0\rangle\, .
\en
Here, $\Delta$ denotes a  
composite field operator made up from the elementary fields
$\Psi$ and $\Phi$, which has the quantum numbers of
the $\chi$-particle. Zero total momentum has been projected out by
summing up over all spatial locations. Further, $J(0)$ stands
for the ``current operator'' defined through the Lagrangian
written down in terms of the elementary fields $\Psi,\Phi$ as
 ${\cal L}^v(0)={\cal L}(0) +vJ(0)+O(v^2)$. An explicit form of the
operators $\Delta,J$ will be inessential in the following.
If Euclidean times $t',t$ become asymptotically large $t'\to +\infty$ and
$t\to -\infty$, the three-point
function $G(t',t)$ goes into
\eq\label{eq:Gas}
G(t',t)\rightarrow \langle 0|\Delta(0)|G\rangle \text{e}^{-E_G(L)|t'|}
\langle G|J(0)|G\rangle \text{e}^{-E_G(L)|t|}\langle G|\Delta^\dagger(0)|0\rangle\, ,
\en
where $|G\rangle$ denotes the ground state with the energy $E_G(L)$ 
(we explicitly indicate the dependence on the volume $L$).

On the other hand, the two-point function of the field $\Delta$ at 
asymptotically large 
Euclidean times behaves as
\eq\label{eq:Das}
D(t',t)=\langle 0|\Delta(t')\Delta^\dagger(t)|0\rangle
\to 
\langle 0|\Delta(0)|G\rangle \text{e}^{-E_G(L)|t'|-E_G(L)|t|}
\langle G|\Delta^\dagger(0)|0\rangle\, .
\en
Extracting the normalization factor 
$\langle 0|\Delta(0)|G\rangle\langle G|\Delta^\dagger(0)|0\rangle$
from Eq.~(\ref{eq:Das}) and using this result in Eq.~(\ref{eq:Gas}), 
one finally determines
the matrix element of the current in the ground state
$ \langle G|J(0)|G\rangle$. In principle, 
it is possible to extract the matrix
elements between the excited states as well.

The problem consists in the following. As was already mentioned, a resonance
does not correspond to any fixed energy level. Then, it is not clear, what is
the relation of the matrix element $\langle G|J(0)|G\rangle$ 
(or, alternatively, of the matrix element between the excited states) to 
the quantity in the r.h.s. of Eq.(\ref{eq:poleshift}). After all, this quantity
is complex whereas all above matrix elements, extracted from the Euclidean
propagators, are real.

In the solution of the above problem, we follow the same path as in the
determination of the pole shift in the external field. The three-point function
(\ref{eq:G}) in a finite volume, 
calculated by using  effective field theory,
is determined by the diagrams shown in Fig.~\ref{fig:vertex} and the volume
dependence of the two-point function is determined by the bubble
diagram shown in Fig.~\ref{fig:bubblesum}. Consequently, measuring the current
matrix element(s) {\em at different volumes} and translating the volume
dependence into momentum dependence by invoking the relation $L=L_m(p)$, 
as discussed in section~\ref{sec:pstar}, one
may parameterize these matrix element(s) in terms of known functions of
$p$ {\em on the real axis}. The final step consists in the substitution
$p\to p_*$ in these expressions, which yield the complex quantity coinciding
with the r.h.s. of Eq.~(\ref{eq:poleshift}).

Finally, we note that by using the same method, 
it is possible to study the form factors at non-zero momentum transfer.
We, however, prefer to address this question separately.

\section{Conclusions}
\label{sec:concl}

Placing a stable particle in a constant external field leads to a
shift of its mass. To first order in the external field, the shift
is given by the form factor of the particle at zero momentum transfer.
The calculations on the lattice can be carried out straightforwardly.

In case of a resonance, the interpretation of the lattice results is difficult,
because a resonance is not described by an isolated energy level in the energy
spectrum. In particular, the form factor of a resonance even at zero
momentum transfer is a complex quantity, whereas all energy shifts are real.

Using non-relativistic effective field theory technique, we have described the
procedure of the extraction of the resonance form factor (at zero momentum
transfer) from the Euclidean lattice data in 1+1 dimensions. 
This procedure can be regarded
as a generalization of L\"uscher's approach to the form factors of the resonances.
In brief, the method consists of the following steps: In the absence of the external
field, one assumes that the effective range expansion is convergent in the
resonance region. Using this fact, one may analytically continue the cotangent
of the scattering phase into the complex plane and determine the location 
of a resonance pole on the second Riemann sheet. Further, it is shown that 
the form of the 
effective range expansion gets modified in the presence of the external field.
Using the modified expansion, one again performs the analytic
continuation into the complex plane and determines the shift of the pole 
position in the external field. 

The main result of the article consists in the following. It is demonstrated 
that the pole shift {\em in a finite volume} up to exponentially vanishing
terms, is determined by the resonance form factor calculated in the infinite
volume. The key observation, which leads to this result, is that approaching
the resonance pole position $p\to p_*$ 
in the complex plane simultaneously implies
the infinite-volume limit  $\lim_{p\to p_*}\text{Im}\,q=-\infty$.

The resonance form factor can be evaluated by using an alternative technique 
as well. Namely, it is possible to directly calculate the corresponding 
three-point function on
the lattice and study its asymptotic behavior at large Euclidean times.
It has been shown in the present paper that by applying the same method
it is possible to construct the resonance form factor at zero momentum
transfer from of the current matrix elements between eigenstates of a 
Hamiltonian in a finite volume that are determined by measuring the 
three-point function.

Finally, it should be stressed that in the case of 3+1 dimensions,
so-called {\em finite fixed points} may exist. These fixed points obey the condition
 $\lim_{p\to p_*}|q|<\infty$ and thus invalidate the proof, given in
section~\ref{sec:pstar}. It remains to be clarified, whether the proof can be
adapted to this case as well. We plan to address this issue in future publications.

\bigskip

{\em Acknowledgments.} We thank J. Gasser, J. Gegelia, M. G\"ockeler, H.-W. Hammer,
T. Hemmert, F.~Niedermaier, V. Pascalutsa,
A. Sch\"afer, G. Schierholz, C. Urbach
 and M. Vanderhaeghen for interesting discussions.

\clearpage

\renewcommand{\thefigure}{\thesection.\arabic{figure}}
\renewcommand{\thetable}{\thesection.\arabic{table}}
\renewcommand{\theequation}{\thesection.\arabic{equation}}

\appendix

\setcounter{equation}{0}
\setcounter{figure}{0}
\setcounter{table}{0}

\section{The loop function}
\label{app:Z00}

To calculate the function $J(p)$ displayed in Eq.~(\ref{eq:Jp}), 
we first perform the integration over the variable $k^0$ using 
Cauchy's theorem. The result in the center-of-mass frame is given by
\eq\label{eq:Jp1}
 J(p)=\int\frac{d^dk}{(2\pi)^d}\frac{1}{2w_\Psi(k)2w_\Phi(k)}\,
\frac{1}{\left(w_\Psi(k)+w_\Phi(k)-P^0\right)}\, ,\quad d=D-1\, .
\en
The integrand in the above equation can be identically 
rewritten in the following form:

\eq\label{eq:identity}
\frac{1}{2w_\Psi(k)2w_\Phi(k)}\,
\frac{1}{\left(w_\Psi(k)+w_\Phi(k)-P^0\right)}
&=&\frac{1}{2P^0}\,\frac{1}{k^2-p^2}
\nonumber\\[2mm]
&+&\frac{1}{2w_\Psi(k)2w_\Phi(k)}\,
\frac{1}{\left(w_\Psi(k)+w_\Phi(k)+P^0\right)}
\nonumber\\[2mm]
&+&\frac{1}{2w_\Psi(k)2w_\Phi(k)}\,
\frac{1}{\left(w_\Psi(k)-w_\Phi(k)-P^0\right)}
\nonumber\\[2mm]
&+&\frac{1}{2w_\Psi(k)2w_\Phi(k)}\,
\frac{1}{\left(w_\Phi(k)-w_\Psi(k)-P^0\right)}\, .
\en
The calculation of the integral given by Eq.~(\ref{eq:Jp1}) proceeds as 
follows~\cite{CGKR}. One first expands the integrand in powers of momenta, 
integrates the result using the dimensional regularization and sums up the 
resulting series.
Doing this, the last three terms on the r.h.s. of
 Eq.~(\ref{eq:identity}) vanish after 
integration, because the integrands turn into polynomials in the momenta.
The non-vanishing result, given in Eq.~(\ref{eq:Jp}), 
is obtained by the integration of the first term.

The calculation of the loop function in a finite volume follows a similar
path. We again perform the integration over the variable $k^0$ and
use the identity Eq.~(\ref{eq:identity}). The sum over the momenta $k$ in the last
three terms (non-singular) exponentially vanishes for large $L$
(see, e.g., Ref.~\cite{sachrajda})
and can  therefore be omitted. Hence, up to these terms, the quantity
$J^L(p)$ from Eq.~(\ref{eq:JL}) is given by
\eq
J^L(p)=\frac{1}{2P^0L}\,\sum_k\frac{1}{k^2-p^2}=
\frac{1}{2P^0L}\,\int_{-\infty}^\infty\frac{dl}{l^2-p^2}\,
\sum_{n=-\infty}^{\infty}\delta\biggl(l-\frac{2\pi}{L}\,n\biggr)\, .
\en
We calculate this expression first below threshold, $p=i\kappa$.
Using the Poisson identity 
\eq \sum_{n=-\infty}^{\infty}\delta(x-n)
=\sum_{n=-\infty}^{\infty}\text{e}^{2\pi i xn}
\en
and integrating over the variable
$l$ with the help of Cauchy's theorem, we get
\eq\label{eq:series}
J^L(p)=\frac{1}{4P^0\kappa}\,\biggl(2\sum_{n=0}^{\infty}\text{e}^{-n\kappa L}
-1\biggr)\, .
\en
Summing up the geometric series in Eq.~(\ref{eq:series}) and continuing 
analytically to the real values of $p$ by substituting $\kappa=-ip$, we finally
obtain the expression given in Eq.~(\ref{eq:Z00}). This expression is nothing
but  L\"uscher's zeta-function in 1+1 dimensions.

\end{document}